\documentclass{optica-article}

\journal{opticajournal} 

\articletype{Research Article}

\usepackage{lineno}

\usepackage{braket}  

\begin{document}

\title{Looking Beyond Visibility: Resolving the Temporal Structure of Continuous-Wave Four-Photon Hong-Ou-Mandel Interference}

\author{Sabine Häussler,\authormark{1,2,*} Meritxell Cabrejo-Ponce,\authormark{1} Ana Maria Torres,\authormark{1,3} Baghdasar Baghdasaryan,\authormark{2} Karen Lozano-Mendéz,\authormark{1} and Fabian Steinlechner\authormark{1,2}}
\address{\authormark{1}Fraunhofer Institute for Applied Optics and Precision Engineering, 07745 Jena, Germany\\
\authormark{2}Friedrich Schiller University Jena, Institute of Applied Physics, Abbe Center of Photonics, 07745 Jena, Germany\\
\authormark{3}Heriot-Watt University Edinburgh Campus, Edinburgh EH14 4AS, United Kingdom}

\email{\authormark{*}sabine.haeussler@uni-jena.de} 

\begin{abstract*} 
Hong-Ou-Mandel interference underpins many photonic quantum information protocols and is commonly characterized by a single visibility parameter. This quantity is typically obtained by averaging over a fixed coincidence window, thereby discarding details of the underlying temporal structure that are critical in continuous-wave processes. Four-photon interference with continuous-wave sources furthermore requires photon pairs with long coherence times, which we generate in microring resonators and measure with ultrafast time-resolved detection. Here, we model and resolve their interference and temporal structure by recording all four photon arrival times and mapping their relative delays onto a three-dimensional coincidence histogram relative to a trigger detection. The resulting structure preserves relative arrival-time correlations that are usually discarded and enables post-processed optimization of coincidence windows to trade off visibility against count rate. This method provides a general benchmark for continuous-wave sources in interference-based quantum network protocols.
\end{abstract*}


\section{Introduction}
\label{sec:introduction}
Hong-Ou-Mandel (HOM) interference~\cite{Ou1987, branczyk2024} is central to applications such as entanglement certification~\cite{graffitti2018}, Gaussian boson sampling~\cite{tillmann2013} and quantum optical coherence tomography~\cite{nasr2003, ibarra-borja2020}. HOM interference of independent sources underpins protocols such as measurement-device-independent quantum key distribution~\cite{lo2012}, twin-field quantum key distribution~\cite{lucamarini2018}, and entanglement swapping~\cite{zukowski1993, pan1998, sun2017, halder2007, samara2021, tsujimoto2018, wu2026a}, where protocol performance directly depends on the indistinguishability of sources. Such protocols can be implemented in two regimes that differ substantially in their spatiotemporal characteristics: while pulsed implementations require stringent optical synchronization across network nodes~\cite{aboussouan2010, kaltenbaek2009a}, continuous-wave (CW) operation can shift synchronization to the electronic domain by exploiting adjustable coincidence windows in post-processing~\cite{halder2007, samara2021, tsujimoto2018, spiess2023,baghdasaryan2025}.

CW HOM interference between independent sources has been demonstrated in Refs.~\cite{halder2008, samara2019, tsujimoto2017, craddock2026a, wu2026a}, but the details of the temporal postselection strategies used in these works remain vague or unspecified. Windowing schemes for extracting HOM dips, coincidence criteria, and data processing methods lack standardized approaches and are rarely discussed at levels of detail required for cross-experimental comparison. This is particularly problematic in CW systems, where visibility and count-rate values require temporal filtering conditions to be specified.

In this work, we present a systematic experimental and post-processing method for CW HOM interference between independent sources. We leverage narrowband photon-pair sources based on spontaneous four wave mixing (SFWM) in microring resonators (MRR) ~\cite{khodadadkashi2025, tagliavacche2025, samara2019}, whose long coherence times enable heralded temporally resolved interference in the CW regime~\cite{samara2019}. Recording the arrival times of all four photons yields three independent relative delays with respect to a trigger photon; four-fold events are thus conveniently represented in a three-dimensional (3D) delay space. This representation makes the characteristic temporal correlations directly visible. Unlike pulsed implementations that rely on optical delay lines, the CW approach enables controlled optimization of delay alignment and coincidence windows without hardware changes. We provide practical guidelines to extract the typical 1D HOM curve. Complementing the experiment, we present a model that accurately describes the observed four-photon interference patterns. 

The manuscript is structured as follows: In Sec.~\ref{sec:theory} we introduce the theoretical background of four-fold HOM interference and develop a simple simulation model. Sec.~\ref{sec:methods} details the post-processing methodology, covering the construction of 3D histograms, their projection onto 1D HOM curves, and the extraction of visibility and count rate parameters. Sec.~\ref{sec:setup} describes the experimental setup, and in Sec.~\ref{sec:results} we discuss the experimental results alongside comparisons with the simulated data. Sec.~\ref{sec:conclusion} summarizes main findings.

\section{Theoretical Background} \label{sec:theory}
Here, we first outline the theory behind the four-fold HOM interference of photons in CW regime. We discuss the physical meaning of various delays and how the four-fold distribution can be modeled in an ideal experiment. 

\subsection{Four-Fold HOM Interference Principles}
\begin{figure}[ht!]
    \centering
    \includegraphics[width=0.65\linewidth]{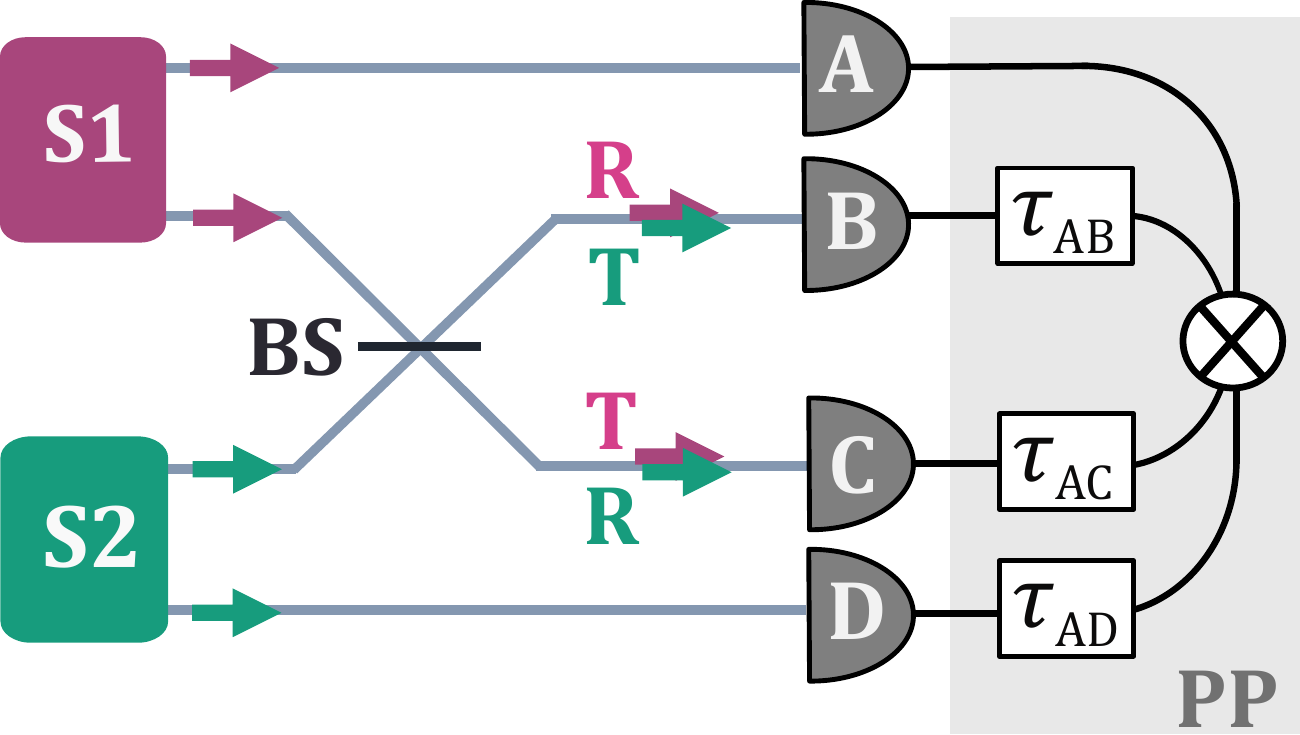}
    \caption{Concept of asynchronous four-fold HOM interference with two long-coherence photon-pair sources: idler photons from sources S1 (magenta) and S2 (green) interfere at a beam splitter (BS), where they can be transmitted (T) or reflected (R). The temporal delays $\tau_\text{AB}$, $\tau_\text{AC}$ and $\tau_\text{AD}$ of the optical channels can be analyzed after detection, in post-processing (PP), instead of physically delaying the photons before the BS.}
    \label{fig:delay_schematic}   
\end{figure}

In a four-fold HOM interference experiment, correlated photon pairs of signal and idler photons are generated from two independent sources, depicted as S1 (magenta) and S2 (green) in Fig.~\ref{fig:delay_schematic}. We will use the following nomenclature throughout the text. The signal photons are used as heralding events in detectors A and D. They provide a temporal reference and herald the corresponding idler photons in detectors B and C. Before detection, the idler photons are directed onto a 50:50 beam splitter with equal reflection and transmission amplitudes. The resulting detection outcomes follow directly from the well-known unitary transformation of creation operators at a beam splitter~\cite{ou2007}. The transformation is applied to the two-photon input state, where the photons originate from sources 1 and 2 and are labeled by their respective arrival times $t_1$ and $t_2$:

\begin{equation} \label{eq:BS}
\hat{b}_{t_\text{1}}^{\dagger}\hat{c}_{t_\text{2}}^{\dagger}|0\rangle \xrightarrow{\text{BS}} 
\frac{1}{2}(\underbrace{\hat{b}'^{\dagger}_{t_1}\hat{b}'^{\dagger}_{t_2}}_\text{RT} - \underbrace{\hat{b}'^{\dagger}_{t_1}\hat{c}'^{\dagger}_{t_2}}_\text{RR}+ \underbrace{\hat{c}'^{\dagger}_{t_1}\hat{b}'^{\dagger}_{t_2}}_\text{TT}-\underbrace{\hat{c}'^{\dagger}_{t_1}\hat{c}'^{\dagger}_{t_2}}_\text{TR} )|0\rangle.
\end{equation}
Here, $\hat{b}^{\dagger}$ and $\hat{c}^{\dagger}$ denote the creation operators at the input modes of the splitter, while $\hat{b}'^{\dagger}$ and $\hat{c}'^{\dagger}$ denote those at the output modes, detected at $B$ and $C$, respectively. The four terms describe the processes in which both photons are transmitted (TT), both are reflected (RR), or one photon is transmitted while the other is reflected (RT, TR).

Scanning the time delay at the beam splitter $\tau_\text{BS} \equiv t_2 - t_1$ controls the temporal indistinguishability of the detected photons, and the resulting HOM curve has a characteristic dip shape. At large delays ($\tau_\text{BS} \to \infty$), the two idler photons arrive at the beam splitter at distinguishable times and no two-photon interference occurs. Here, TT and RR events can be detected as four-fold coincidences, while TR and RT events remain undetectable with click detectors (Fig.~\ref{fig:timetags}). As $\tau_\text{BS}$ approaches zero, the TT and RR amplitudes become indistinguishable and cancel by HOM interference. The rate of coincidences in modes $b$ and $c$ therefore reaches its minimum at $\tau_\text{BS} = 0$, producing the characteristic HOM dip. 

The HOM dip visibility quantifies photon indistinguishability by comparing the four-fold coincidence rate at the dip to a reference rate for non-interfering coincidences~\cite{branczyk2024, ou2007}:
\begin{equation} \label{eq:vis}
V_{\text{HOM}} = 1 - \frac{R_0}{R_\infty},
\end{equation}
where $R_0 = R(\tau_\text{BS} = 0)$ is the coincidence rate for a defined coincidence window at the center of the interference dip and $R_\infty=R(\tau_\text{BS}\to \infty)$ is the coincidence rate at large delays, where the photon pairs are fully distinguishable.

A key aspect of this scheme is the temporal overlap of interfering photons. In pulsed experiments, photons' temporal overlap, and hence their interference, is controlled via physical optical delay lines introduced before the beam splitter, whereas a well-defined temporal mode is ensured by quantum state engineering techniques. For CW processes, by contrast, the relevant relative delays can be selected in post-processing from the recorded detection times. When the detection timing jitter $\tau_\text{jitter}$ is shorter than the coherence time $\tau_\text{coh}$ of the photon pairs \cite{Huang2010,baghdasaryan2025}, time-resolved detection effectively projects the partner photon onto a single temporal mode. This allows the photon-pair sources to operate asynchronously, without any external clock defining their emission times. 

Analysis of the relative detection times further allows the contributions of non-overlapping photon pairs to be distinguished according to whether both photons are transmitted (TT) or reflected (RR) (see Fig.~\ref{fig:timetags}a). Separating these contributions enables their independent analysis and thus provides a more detailed characterization of the interference pattern. Away from the HOM dip, the arrival times of the heralding photons provide which-path information and interference is suppressed. HOM interference occurs only for detection events where the RR and TT paths are temporally indistinguishable and there is no such which-path information available \cite{pittman1996}.

\subsection{Modeling Four-Folds Distribution} \label{sec:model}
\begin{figure}[ht]
    \centering
    \includegraphics[width=0.95\linewidth]{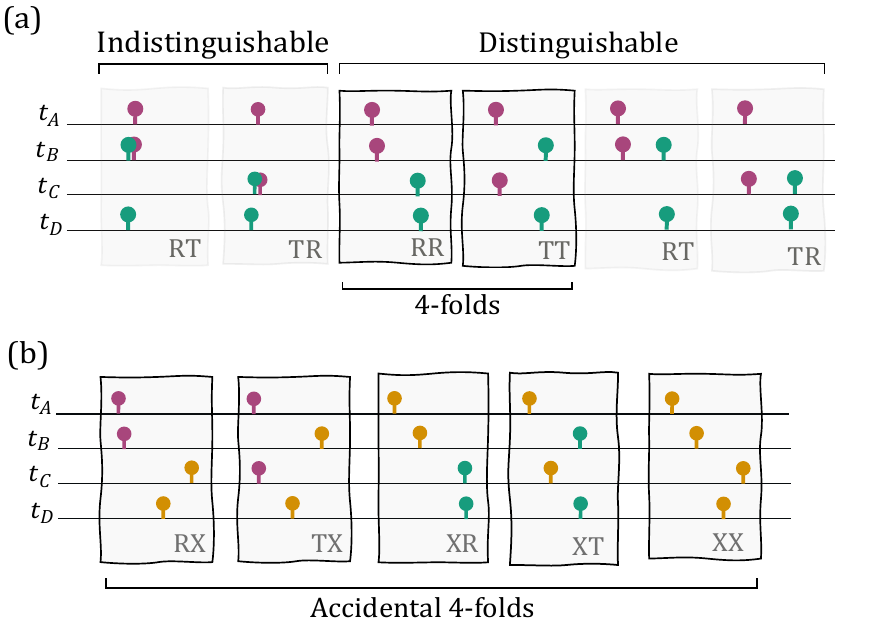}
    \caption{Distribution of possible photon detections in a four-fold HOM experiment: photon pairs generated by two independent sources (magenta and green) arrive at the detection channels with temporal correlations. We distinguish between cases in which the four-fold events originate exclusively from one photon pair from each source (a), and cases in which they arise from uncorrelated photons (b). In (a), we label these events as reflected-transmitted (RT), transmitted-reflected (TR), reflected-reflected (RR) and transmitted-transmitted (TT) photon events. With temporal indistinguishabilty, the TR and RT cases cannot be resolved as two-photon-pair events when using click detectors. In addition, the RR and TT contributions interfere destructively and cancel out. Only temporally distinguishable RR and TT events can be detected as four-fold coincidences. In (b), accidental coincidence events arise from background photons (orange) that are detected at random times. We label these events as: reflected-accidental (RX), transmitted-accidental (TX), accidental-reflected (XR), accidental-transmitted (XT) and accidental-accidental (XX).}
    \label{fig:timetags}
\end{figure}

To model the four-fold coincidence rates as a function of the relative time delays between the four detection channels, we first consider the ideal case without accidental coincidence events from multi-pair emission and background noise counts. Fig.~\ref{fig:timetags} illustrates the probability amplitudes leading to detection of four photons in a four-fold HOM experiment. Events in which both photons exit the beam splitter through a common output port do not lead to a four-fold coincidence detection. When the relative delay is much larger than the coherence time, no HOM interference is observed. In this regime, the four-fold coincidence events arise from two distinguishable contributions: both photons being transmitted (TT) or both being reflected (RR) at the beam splitter (Fig.~\ref{fig:delay_schematic}).

In the case of RR events, a photon detection in Channel A at time $t_\text{A}$ heralds the presence of its pair in Channel B, detected at time $t_\text{B}$, and similarly for the photon pair in channels C and D. For the case of TT events, a detection in Channel A heralds its pair in Channel C, while Channel D heralds a corresponding event in Channel B. The distribution of time delays between photons of the same pair follows their corresponding temporal coherence function $G^{(2)}(\tau)$ (Eq.~\eqref{eq:G2}). In this work, we focus on MRRs, for which this function is ideally Lorentzian in frequency, corresponding to a double-exponential decay in time with characteristic decay time $1/\gamma$ (see Appendix \ref{app:Theory_Biphoton_mode_fnc}).

For direct comparison with the experiment, we work with photon-flux amplitudes rather than normalized probability amplitudes. To this end, we define the photon-flux amplitude $\mathcal{A}$ on a discrete three-dimensional delay grid $(\tau_{AB},\tau_{AC},\tau_{AD})$, where we take Channel A as the time reference and discretize the relative delays
\begin{equation}
\tau_\text{AB}= t_\text{B}-t_\text{A}, \quad \tau_\text{AC}= t_\text{C}-t_\text{A}, \quad \tau_\text{AD} = t_\text{D}-t_\text{A}
\end{equation}
into 3D time bins of width $\delta\tau$. The squared magnitude $|\mathcal{A}|^2$ directly yields the four-fold coincidence rate $r$ per bin. This formulation allows direct comparison between the model and the measured four-fold coincidence histogram without additional rescaling.

Source S1 produces a photon pair in channels A–B. Conditioned on a detection event in Channel A at time $t_\text{A}$, the amplitude of the heralded photon in Channel B decays as $e^{-\frac{\gamma_1}{2}|t_\text{B}-t_\text{A}|}=e^{-\frac{\gamma_1}{2}|\tau_\text{AB}|} $, reflecting the Lorentzian linewidth $\gamma_1$ of source S1. Source S2 independently produces a pair in channels C–D with an analogous factor $e^{-\frac{\gamma_2}{2}|t_\text{C}-t_\text{D}|}= e^{-\frac{\gamma_2}{2}|\tau_\text{AC}-\tau_\text{AD}|}$ (see Appendix \ref{app:Theory_Biphoton_mode_fnc}). For the RR path, the two-source amplitude factorizes as

\begin{equation}
      \mathcal{A}_{RR}(\tau_\text{AB}, \tau_\text{AC}, \tau_\text{AD})=\mathcal{A}_\text{0,RR} \, e^{-\frac{\gamma_1}{2}|\tau_\text{AB}|}\,
  e^{-\frac{\gamma_2}{2}|\tau_\text{AC}-\tau_\text{AD}|},
\end{equation}
where $\mathcal{A}_\text{0,RR}$ is a constant normalization factor. For the TT path, the roles of B and C are exchanged: the photon pair from S1 occupies channels A–C, while the pair from S2 occupies channels B–D, giving:
\begin{equation}
     \mathcal{A}_\text{TT}(\tau_\text{AB}, \tau_\text{AC}, \tau_\text{AD})=\mathcal{A}_\text{0,TT}\, e^{-\frac{\gamma_1}{2}|\tau_{AC}|}
  \,e^{-\frac{\gamma_2}{2}|\tau_{AB}-\tau_{AD}|}.
\end{equation}

For any relative temporal delay, the four-fold coincidence rate depends only on the two contributions  RR and TT (Eq.~\eqref{eq:BS}) and is given by
\begin{equation}
    r_\text{HOM}(\tau_\text{AB}, \tau_\text{AC}, \tau_\text{AD})
 = \bigl|\mathcal{A}_\text{TT}-\mathcal{A}_\text{RR}\bigr|^2.
\end{equation}
The relative minus sign between the RR and TT amplitudes follows directly from the unitary beam-splitter transformation in Eq. \eqref{eq:BS}. At zero delay, $\tau_\text{AB}=\tau_\text{AC}=\tau_\text{AD}=0$, these two contributions cancel. Hence, under ideal conditions, such as a balanced beam splitter and perfectly indistinguishable amplitudes, the coincidence rate vanishes due to destructive interference.

The prefactors $\mathcal{A}_\text{0,RR}$ and  $\mathcal{A}_\text{0,TT}$ carry units $[s^{-1/2}]$, ensuring $r_\text{HOM}$ has units $[\text{s}^{-1}]$. To estimate their absolute values, we assume Poissonian statistics, as commonly done to estimate coincidences between uncorrelated events \cite{janossy_rate_1944,Neumann2021}. Therefore, for uncorrelated single-count rates $R_i$ and $R_j$ in two distinct channels, the coincidence rate within a time window $\Delta\tau$ is $R_iR_j\Delta\tau$. We extend this argument to four-fold events by treating the two detected photon pairs as independent. Let $R_{kl}$ denote the two-fold coincidence rate between channels $k$ and $l$ per histogram bin of width $\delta\tau$, rather than the rate integrated over the full two-fold histogram. The rate at which a photon pair in channels A and B is found in the same bin as an independent photon pair in channels D and C is then $R_\mathrm{AB}R_\mathrm{DC}\delta\tau$. We therefore define the normalization factor as:
\begin{equation}
    \mathcal{A}_\text{0,RR}^ 2=R_\text{AB} R_\text{DC} \delta\tau
\end{equation}
for the RR events. Accordingly, for the TT events:
\begin{equation}
    \mathcal{A}_\text{0,TT}^2 = R_\text{AC} R_\text{DB} \delta\tau.
\end{equation}

We obtain $R_\text{AB}, R_\text{AC}, R_\text{DB}, R_\text{DC}$ and $\gamma_1$, $\gamma_2$ from a double exponential fit function of the measured two-fold histograms, where we assume perfect Lorentzian spectrum (see Appendix~\ref{app:fit}).

\subsection{Modeling of Accidental Four-Folds}
\label{sec:Acc}
Accidental coincidences require particular care in CW experiments because their rate depends explicitly on the postprocessing strategy. For the discussion of accidental counts it is useful to distinguish genuine four-photon detection events, in which both sources emit a correlated photon pair and the two interfering photons overlap at the beam splitter, from accidental four-fold coincidences stemming from uncorrelated detections. Note that only the former can exhibit the two-photon interference required, for example, for entanglement swapping.

The leading order of  four-fold accidentals consists of one correlated photon pair accompanied by two uncorrelated detections within the same coincidence window (Fig.~\ref{fig:timetags}(b)). These uncorrelated detections may originate from detector dark counts, leakage of the pump laser driving the SFWM process, photons generated in neighboring frequency bins, or photons whose partners were lost.

Consider, for example, a reflected-accidental (RX) event, where channels A and B detect a correlated pair, while channels C and D detect independent background photons. The resulting timing-distribution follows the coherence function for the A-B pair but is independent of the detection times $t_\text{C}$ and $t_\text{D}$. Analogous contributions arise in cases where the correlated pair is detected in other channel combinations: A-C for a transmitted-accidental (TX) event, D-C for an accidental-reflected (XR) event and D-B for an accidental-transmitted (XT) event (Fig.~\ref{fig:timetags}b). 

We can therefore define the accidental contribution of all these events with one correlated pair and two uncorrelated detections. These four leading-order accidental contributions per bin are then:
\begin{align} \label{eq:acc}
r_\text{RX} ={R_\text{AB}   R_\text{C}  R_\text{D}\delta\tau ^2}  e^{-\gamma_1|\tau_\text{AB}|} \\
r_\text{TX} ={R_\text{AC} R_\text{B}  R_\text{D}  \delta\tau ^2}e^{-\gamma_1|\tau_\text{AC}|}\\
r_\text{XR} ={R_\text{DC}R_\text{B}  R_\text{A} \delta\tau ^2}e^{-\gamma_2|\tau_\text{AC}-\tau_\text{AD}|}\\
r_\text{XT} ={R_\text{DB}R_\text{C}  R_\text{A} \delta\tau ^2}  e^{-\gamma_2|\tau_\text{AB}-\tau_\text{AD}|},
\end{align}
where $R_i$ denotes the single-count rate in channel $i$, and $R_{kl}$ the
two-fold coincidence rate between channels $k$ and $l$ per histogram bin of
width $\delta\tau$.

For clarity, the temporal distribution of these different noise contributions is shown in Appendix~\ref{app:noise}.

The full model of the distribution of four-fold accidental rates is then:
\begin{equation}
    r_{\text{acc}} =r_\text{RX}+r_\text{TX}+r_\text{XT}+r_\text{XR},
    \label{eq:model_noise}
\end{equation} 
where we have omitted the contribution  $r_\text{XX}= R_\text{A}  R_\text{B}    R_\text{C}  R_\text{D} \delta\tau ^3$, which corresponds to the case in which all four channels detect independent photons and is usually negligibly small.
Combining these unavoidable accidental four-folds with true four-folds (meaning two pairs of correlated photon pairs) gives the full picture of detected four-fold events:
\begin{equation}
    r(\tau_\text{AB}, \tau_\text{AC}, \tau_\text{AD} ) =r_\text{HOM}+r_\text{acc},
\end{equation} 
which is a 3D histogram of four-fold coincidence rates with Channel A as the time reference. This method allows us to model the HOM curve and, importantly, to accurately estimate the total four-fold count rate, accidental rate, and expected visibility under ideal conditions, using only single count rates and parameters of the two-fold coincidence histograms of the two sources, even before conducting the full four-photon experiment.

\section{Post-Processing Methods} \label{sec:methods}
\label{sec:postprocessing}

To analyze the four-fold coincidence distribution, we first align the four channels in time with respect to each other. From this data we extract the four-fold events and map them into a 3D histogram, representing the distribution over the three relative time delays defined with respect to a reference channel. Then the HOM curve can be extracted from that 3D histogram, for which we discuss two approaches: the conventional single-channel-delay method and an improved two-channel-delay method. These methods are explained here using simulated data. In Sec.~\ref{sec:results} it follows a discussion in which these methods are applied to experimental data.

\subsection{Temporal Channel Calibration}
Differences in optical path length, detector latency, and electronic delay introduce timing offsets between the four detection channels. In a first step, offsets relative to a temporal reference are removed from the raw time-tag data. 

Channel A, with time tags at $t_\text{A}$, is used as the temporal reference and remains unchanged. The offsets of Channels B and C are determined relative to Channel A from the center of their respective A--B and A--C two-fold correlation peaks, obtained from detections of photon pairs generated by the same source.

The offset of Channel D is determined with the delay from the two-fold correlation peak with Channel B, and must also include the delay of Channel B itself, to reference it with respect to A. This calibration removes optical and electronical offsets to ensure that there is no additional delay in the two path alternatives (TT and RR) within their coherence time. 

\subsection{Temporal 3D Mapping}
\begin{figure}[ht]
    \centering
    \includegraphics[width=0.99\linewidth]{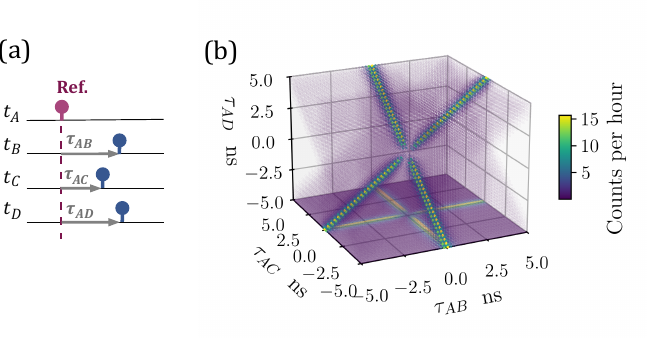}
    \caption{Temporal 3D mapping of the modeled four-fold coincidence distribution. (a) Visualization of the relative temporal delays $\tau_\text{AB}= t_\text{B}-t_\text{A}$, $\tau_\text{AC}= t_\text{C}-t_\text{A}$, $\tau_\text{AD} = t_\text{D}-t_\text{A}$ as defined across the four time tagging channels. (b) Simulated 3D histogram. Here, the four-fold events are distributed along two diagonal branches, corresponding to the correlation behaviour of the RR and TT events.}
    \label{fig:3d}
\end{figure}

A four-fold coincidence event is recorded when all four channels (A, B, C, D) register a detection within a short time of each other, where Channel A serves as the trigger and channels B, C, and D as stop channels. To investigate the temporal structure of these events, we construct a 3D data set, where the three axes represent the relative delays: $\tau_\text{AB}= t_\text{B}-t_\text{A}$, $\tau_\text{AC}= t_\text{C}-t_\text{A}$, $\tau_\text{AD} = t_\text{D}-t_\text{A}$ (Fig.~\ref{fig:3d}a). We discretize the data set of this 3D histogram by the bin width $\delta\tau=50$\,ps, ensuring the overall resolution is not limited by the timing jitter. 

The data set of the 3D histogram can be visualized in a scatter plot, with the number of counts per bin indicated by color. For the simulated data, this plot is shown in Fig.~\ref{fig:3d}b. 

Four-fold detections resulting from one photon pair from each source form two diagonal branches. Each branch is defined by the correlations within the two pairs, while the delay between the pairs, given by $\tau_\text{AD}$, is random since the CW sources emit independently. For TT events, channels A and C detect one pair ($\tau_\text{AC}\approx 0$) and channels D and B the other ($\tau_\text{AB}-\tau_\text{AD}\approx 0$), so the events spread along the diagonal $\tau_\text{AB}=\tau_\text{AD}$ in the plane $\tau_\text{AC}=0$. Similarly, RR events occur along $\tau_\text{AC}=\tau_\text{AD}$ in the plane $\tau_\text{AB}=0$. Both branches intersect at $\tau_\text{AD}=0$, where the photons from the two sources become temporally indistinguishable and HOM interference suppresses four-fold coincidences. The dominant contribution to accidental coincidences appear as four planes crossing the diagonal branches of the true four-folds (see Appendix~\ref{app:noise}). 

\subsection{3D Coincidence Window}
The 3D histogram provides complete temporal information about four-fold detections.  To extract a coincidence rate $R$ from this distribution, we define a finite timing window in the relative-delay space. This coincidence window is a rectangular box centered at the position $(\tau_\text{AB},\tau_\text{AC},\tau_\text{AD})$  in delay-space with side lengths $\Delta\tau_\text{AB}$, $\Delta\tau_\text{AC}$ and $\Delta\tau_\text{AD}$. Each of the side lenghts specifies the maximum allowed time difference between detection events that are counted as a coincidence. The four-fold coincidence rate $R(\tau)$ is obtained by summing over all events within this box. The choice of coincidence window involves a  trade-off: The window should be wide enough to capture all events for which quantum interference occurs, but narrow enough to select events that have a high degree of temporal indistinguishability and reduce the relative contribution of accidental coincidences. 

A one-dimensional HOM interference pattern is obtained by translating the center of the box, defined by $\tau_\text{AB}$, $\tau_\text{AC}$, and $\tau_\text{AD}$, through the 3D delay-space and evaluating the coincidence rate at each position. In the following, we consider two trajectories that  can be applied to reveal the typical HOM dip. 

\subsection{Method 1: Single Channel Delay}
\begin{figure}[ht]
    \centering
    \includegraphics[width=\linewidth]{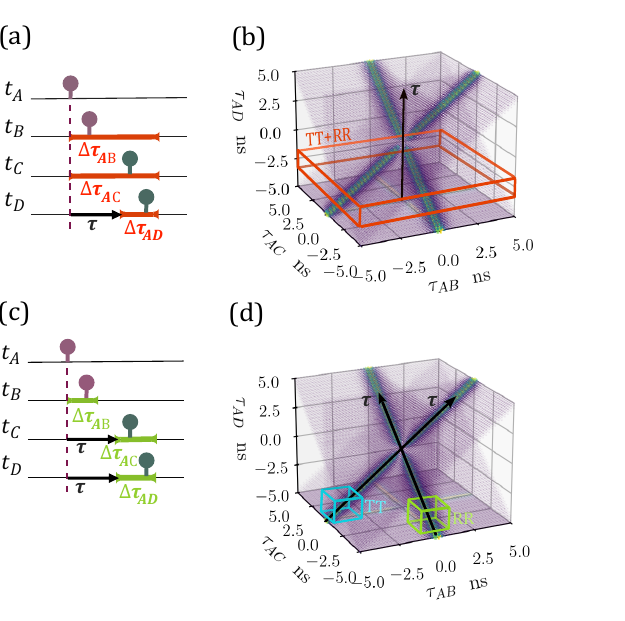}
    \caption{Two different methods to scan HOM interference. Method 1 and Method 2 are depicted in the upper (a, b) and lower (c, d) rows, respectively. The first column (a, c) shows the distributions of detection times together with the trigger detection. The second column (b, d) illustrates the scan of the coincidence window and its corresponding dimensions for each method. In Method 1, the coincidence window (red box) is scanned along a delay $\tau$ (black arrow). In Method 2 (lower row), the coincidence window is shifted in two channels, resulting in a small coincidence window being translated diagonally across the three-dimensional histogram.}
    \label{fig:CompareMethods}
\end{figure}

To obtain the HOM curve, we scan the relative delay between the two heralding channels A and D. Taking Channel A as the temporal reference, we vary the detection time of Channel D while keeping the heralded channels B and C unchanged \cite{baghdasaryan2025}. The effect with regards to the registered coincidence counts is equivalent to that of a physical delay line inserted before detector D. By shifting this delay in post-processing, the timestamp of the heralding detection event is shifted relative to the detection time of its paired photon at the beam splitter. The detections after the beam splitter (B and C) maintain a fixed delay relative to the reference Channel A (Fig.~\ref{fig:CompareMethods}a). In the 3D representation, this corresponds to shifting $\tau_\text{AD}$ while keeping the box centered at $\tau_\text{AC}= \tau_\text{AB}=0$. The box shifts along the $\tau_\text{AD}$ axis as illustrated in Fig.~\ref{fig:CompareMethods}b.

In this method, both TT and RR events contribute to the total coincidence rate $R_\infty=R_\text{RR+TT}(\tau)$. This combined rate serves as the reference, indicating the expected coincidence rate at $\tau=0$ when two-photon interference is absent. 

The method has an important limitation when narrow windows are applied to channels B and C. As the heralding detection time in D is delayed, the coincidence windows in B and C remain fixed. Thus, once $|\tau_\text{AD}|$ exceeds $\Delta\tau_\text{AB}$ or $\Delta\tau_\text{AC}$, respectively, the corresponding events fall outside the coincidence windows and are not counted as valid four-fold coincidences. If the windows are narrower than the coherence time, the four-fold coincidence rate drops to zero before the baseline is reached, making it impossible to determine the distinguishable reference $R_\infty$ (see Sec.~\ref{sec:Compare_methods} and Fig.~\ref{fig:HOM_methods}). Therefore, wide coincidence windows are required for $\Delta\tau_\text{AB}$ and $\Delta\tau_\text{AC}$. This means the method evaluates the HOM visibility of the heralded photons without temporal filtering in modes B and C. However, using these wider windows inevitably increases the contribution of accidental coincidences and reduces the visibility [Eq.~\eqref{eq:vis}].

\subsection{Method 2: Dual-Channel Delay}
To overcome the limitations of the single-channel delay approach, we introduce a second method in which the coincidence windows can remain small. Instead of delaying only one heralding channel, this method shifts both detection channels of the same photon pair together (Fig.~\ref{fig:CompareMethods}c), preserving the temporal correlation within each pair while delaying S2 events relative to S1. For example, to capture RR events, channels D and C are scanned together while B remains at zero delay, so that the large-delay region yields the proper incoherent reference $R_\infty$. In the 3D representation, this corresponds to moving the coincidence box diagonally along the RR branch (increasing $\tau_\text{AD}$ and $\tau_\text{AC}$ together, while keeping $\tau_\text{AB} = 0$), as shown in Fig.~\ref{fig:CompareMethods}d. Analogously, the TT branch is scanned by delaying channels D and B together (fixed $\tau_\text{AC} = 0$).

This method provides the four-fold rate $R_\text{RR}(\tau_\text{AD})$ by delaying D and C together, and the rates $R_\text{TT}(\tau_\text{AD})$ by delaying D and B together. At delays much larger than the coherence time, the photon pairs are distinguishable and each scan isolates only one beam splitter outcome (either the RR or the TT branch), so both must be summed to recover the full non-interfering rate. Consequently, to reconstruct the total count rate of these non-interfering photons, one must sum the coincidences across both possible beam splitter outcomes: 
\begin{equation}
R_\infty = R_\text{RR}(\tau_\text{AD}) + R_\text{TT}(\tau_\text{AD})
\label{eq:count_rate}
\end{equation}
for $|\tau_\text{AD}|\gg \tau_\text{coh}$. 

The key advantage of this method is that both photons of a pair are shifted by the same delay $\tau$, so their time difference at the detectors is unchanged at any delay setting and no events are lost as a result of scanning $\tau$. This enables accurate determination of the HOM curve baseline and $R_\infty$ even with narrow coincidence windows, allowing visibility and count rate to be quantified and optimized for any window size. 

\subsection{Comparing Different Methods for Reconstructing the HOM Dip}
The two methods differ substantially in which channels are temporally filtered. Method 1 applies the delay exclusively to the heralding channels A and D, while the beamsplitter output channels C and D remain fixed. Method 2 shifts the heralding channels and their corresponding beam splitter outputs together. For example photons A and B  and D and C together, for the case of RR events. This directly affects the minimum usable coincidence window and how the reference rate $R_\infty$ is constructed.

In the CW regime, temporal indistinguishability is selected through the post-processing conditions, unlike in pre-synchronized pulsed experiments. Consequently, the two reconstruction methods yield different visibility values, since they correspond to different temporal post-selections of the same data. Comparisons between HOM measurements across different methods or experiments, therefore require a clear definition of the procedure used. In particular, the definitions of $R_0$ and $R_\infty$, and thus the visibility, depend on the method used. In other methods than used here, the visibility may be normalized to either TT or RR branch alone, so that the total reference count rate can differ by a factor of two.

Appendix~\ref{app:HOM_method_3} describes a third method that uses both heralding detections as references. This separates the delay between the two photon pairs from the delay between the photons within each pair, which can simplify the analysis of coincidence-window effects. However, this separation applies only when considering TT or RR events individually.

\subsection{Accidental Coincidences}
In CW operation, accidental coincidences constitute an intrinsic background. Here, we assume that higher-order detection events are statistically independent and temporally uncorrelated, such that their occurrence is described by Poissonian statistics. In the model introduced in Section~\ref{sec:model}, the four dominant accidental contributions appear as four planes in the 3D delay space: two diagonal and two vertical planes (Fig.~\ref{fig:3d}b).
This characteristic geometry allows the accidental  background to be identified and its contribution to the HOM visibility estimated. At the HOM dip center ($\tau_\text{AB} = \tau_\text{AC} = \tau_\text{AD} = 0$), all four planes intersect, whereas along each of the TT and RR branches only two planes contribute.  Although the accidental background is not uniform in the delay space, it contributes equally to $R_0$ and $R_\infty$. It therefore does not change their difference but adds a constant offset that reduces the measured visibility, which can be quantified directly from the plane geometry.
\begin{figure}[ht]
    \centering
    \includegraphics[width=\linewidth]{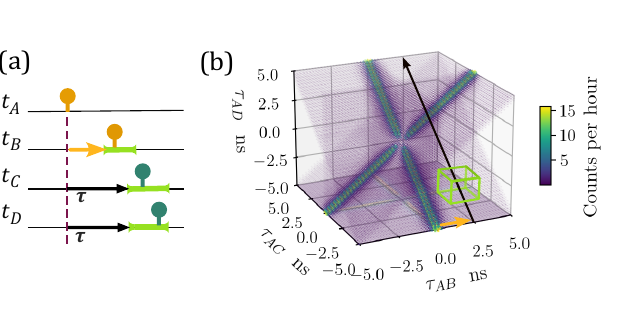}
    \caption{Scheme of the accidental-scanning method, shown here as an example for XR events. One channel (here B) has an offset, while the delay of two other channels (here C and D) are scanned. Consequently, the coincidence window integrates XR-accidentals only. Similar procedures follow for the other accidental contributions. (a) for the delay across the four detection channels, (b) for scanning coincidences in the 3D histogram.}
    \label{fig:scanAcc}
\end{figure}

To quantify the accidental contribution, we introduce delay offsets that shift the coincidence window onto the accidental planes, away from true four-fold events (Fig.\ref{fig:scanAcc}). By subtracting the estimated accidental contribution from the measured count rates, we obtain the corrected values: $R_{\infty,\text{corr}} = R_\infty - R_\text{acc}$ and $R_{0,\text{corr}} = R_0 - R_\text{acc}$, revealing the intrinsic source quality (Fig.\ref{fig:vis}). A corrected visibility [Eq.~\eqref{eq:vis}] can be useful for source characterization, but should not be used when modeling quantum key distribution performance, where this background noise is physically present.

\section{Experimental Setup} \label{sec:setup}
\begin{figure}[htbp!]
    \centering
    \includegraphics[width= \linewidth]{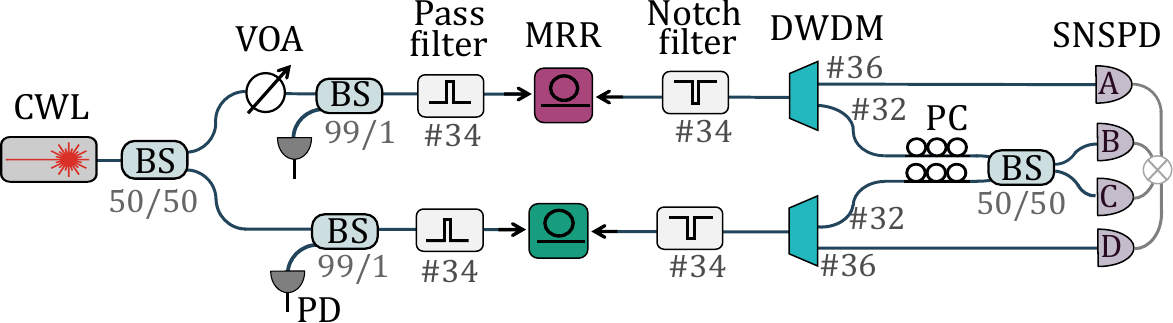}
    \caption{Experimental setup for CW HOM interference. Two microring resonators (MRRs) are used as photon-pair sources, and their idler photons interfere at a beam splitter without an additional physical delay line. CWL: CW laser, BS: beam splitter, VOA: variable optical attenuator, DWDM: dense wavelength division multiplexing filter, SNSPDs: superconduction nanowire single photon detectors, PD: photo detector, PC: polarization controller.}
    \label{fig:setup}
\end{figure}

Our experimental setup is shown in Fig.~\ref{fig:setup}. It consists of two independent silicon nitride chips, each containing a MRR (Ligentec). These resonators generate correlated photon pairs through SFWM and are pumped on resonance by a 1550\,nm CW laser. We amplify the laser using an erbium-doped fiber amplifier (EDFA) to ensure sufficient pump power for photon-pair generation. The pump light is split with a fiber beam splitter, followed by a polarizing beam splitter (PBS) and a dense wavelength division multiplexing (DWDM) filter in each path to clean the polarization and spectrum of the input light. The filtered light is coupled to the chips via lensed fibers with a coupling loss of $\approx$3\,dB per facet. One chip is heated to shift the resonance wavelength to match that of the second MRR. At the output, the pump light is separated from the single photons using a DWDM filter. A second DWDM filter split signal and idler photons into channels 32 and 36 of the ITU (International Telecommunication Union) DWDM grid. The signal photons are used for heralding in channels A and D, while the idler photons are directed to a fiber beam splitter (BS), with output channels B and C. For this, the polarization state of each path is aligned with polarization controllers (PC) to achieve polarization indistinguishability. All channels are connected to  superconducting nanowire single photon detectors (SNSPD) from Single Quantum and the detected time stamps of the single photon detections are processed with time taggers (Swabian Instruments). The detectors feature a detection efficiency in the range from 81-91\;$\%$, dark counts of approximately 200\;cps and a FWHM jitter ranging from 13-17 ps. The timetagger has a digital resolution of 1 ps and a specified FWHM jitter of 8 ps. Together, these result in a jitter budget for two-fold coincidences of approximately 34 ps, which is substantially shorter than the coherence time of the MRR sources.

The two MRRs have a free spectral range of 101\,GHz and coherence times of 433\,ps (445\,ps)  for S1 (S2). The detected photon pair rate, obtained by summing coincidences from both output ports of the BS, is 17\,kpairs/s (19\,kpairs/s), evaluated using a coincidence window of 500 ps. The heralding efficiency, defined as the summed coincidence rate divided by the respective heralding detector singles rate, is 1.9\,\% (5.7\,\%). This corresponds to a coincidence to accidental ratio (CAR) of 36 (95). 
The significant performance difference between the two sources is caused by differences in the filter quality. The experiment is conducted over a duration of 258 minutes.

\section{Experimental Results} \label{sec:results}
We now apply the analysis introduced above to the experimental data obtained with the setup described in Sec.~\ref{sec:setup} and compare the results with the simulated model presented in Sec.~\ref{sec:theory}. Additionally, a discussion on the relevant figures of merit for applications is provided, which focuses on visibility and count rate. 

\subsection{3D Histogram}
\begin{figure}[htbp!]
    \centering
    \includegraphics[width=\linewidth,trim=2 1 1 1,
    clip]{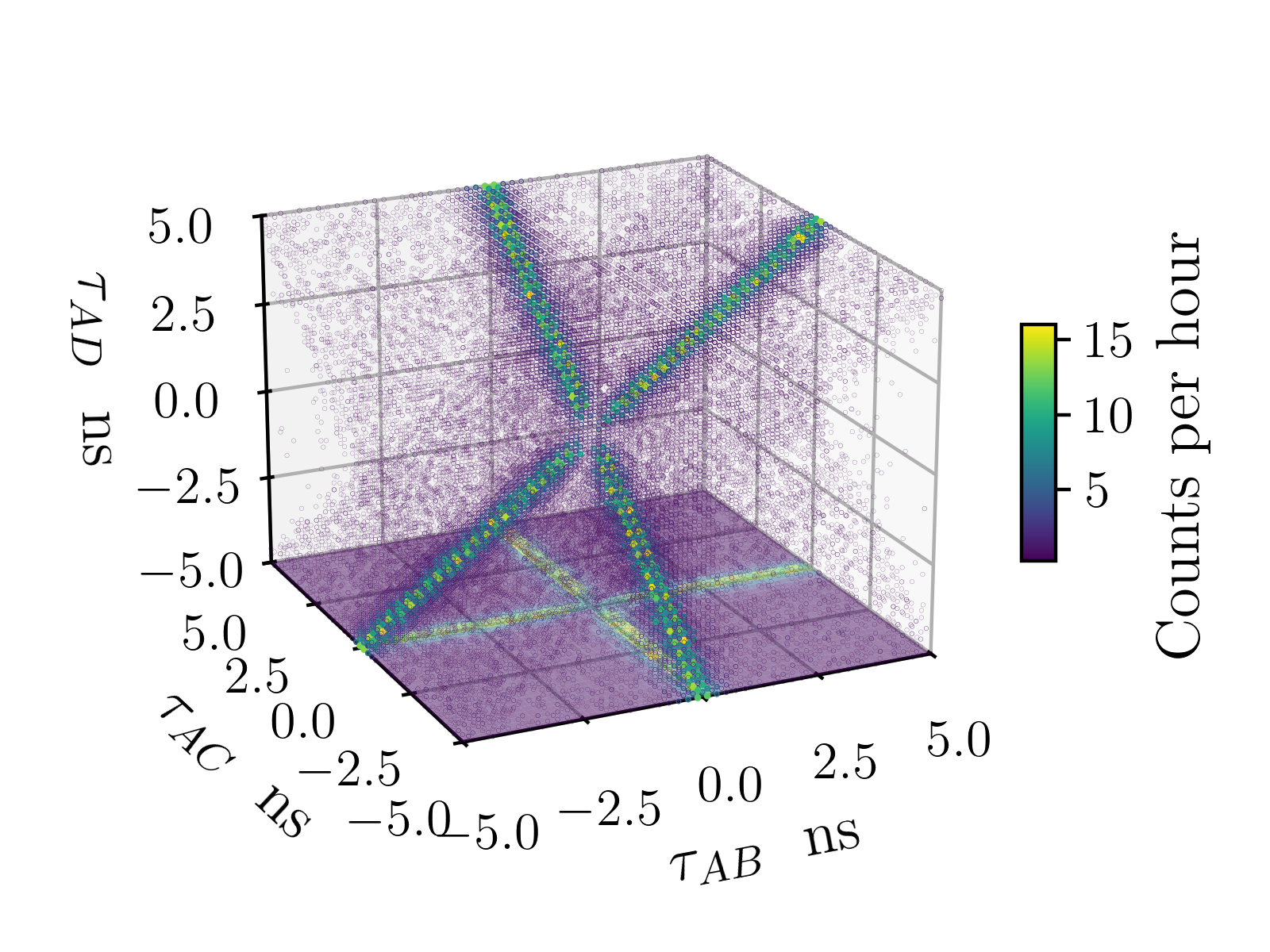}
    \caption{Experimental 3D histogram of four-fold events, using channel~A as the start reference and plotted over the delays relative to the three other channels. The coincidences are distributed along two distinct correlation branches corresponding to TT and RR photon-pair events. At the intersection, the four-fold coincidences drop toward zero. Accidental events are mainly distributed in four planes, as described in Sec.~\ref{sec:Acc} and Appendix~\ref{app:noise}.}
    \label{fig:3dHist}
\end{figure}

We first map the measured four-fold coincidence events onto the 3D delay space histogram introduced previously. The resulting experimental data is shown in Fig.~\ref{fig:3dHist}. 
In agreement with the model, the coincidences are distributed along two distinct correlation branches, corresponding to TT and RR photon-pair events. Each time stamp in the interfering channels B and C is correlated with one of the heralding channels A or D, and the cross-section of each branch corresponds to the respective cross-correlation functions of the sources (Appendix~\ref{app:HOM_method_3}). At the intersection of the two branches, the count rate drops toward zero, as expected from HOM interference.

\subsection{Comparing Different Methods} \label{sec:Compare_methods}
\begin{figure}[ht]
    \centering
    \includegraphics[width=\linewidth]{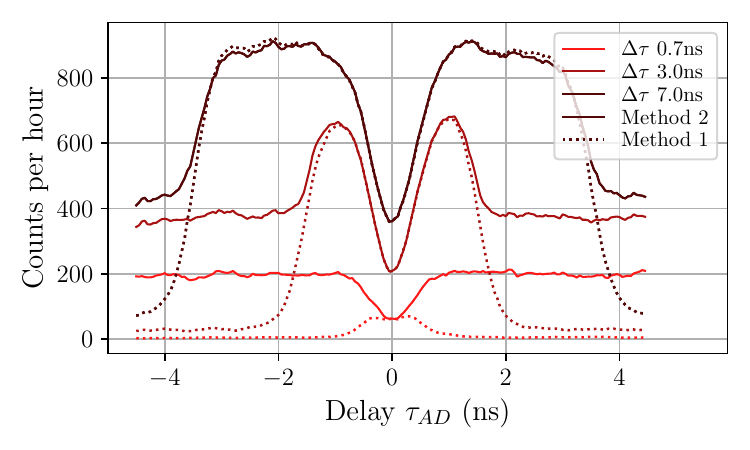}
    \caption{HOM curve obtained with Method 1 (dotted) and Method 2 (solid). For Method 2, only the scan along the TT branch is shown; the RR scan yields equivalent curves. To compare the two different methods, $\Delta\tau_\text{AD}$ is fixed at 0.1ns. For a short window of $\Delta\tau=\Delta\tau_\text{AB}=\Delta\tau_\text{AC}=0.7$\,ns, Method 1 does not produce a usable HOM curve because no baseline forms, whereas Method 2 already shows a well-defined dip. At $\Delta\tau=3.0$\,ns, Method 2 begins to include events from the crossing region of the TT and RR branches, leading to side peaks next to the central dip, while Method 1 starts to form the baseline. For $\Delta\tau_\text{AB}=\Delta\tau_\text{AC}=7$\,ns, Method 1 produces a HOM curve with a broad baseline, but at such long coincidence windows the visibility is reduced. For Method 2, a two-level baseline emerges at long coincidence windows: the inner baseline contains both RR and TT contributions ($R_\text{RR+TT}$), while the outer baseline corresponds to just one branch; here $R_\text{TT}$. For consistency, the sum of the outer baselines of the two branches gives the reference $R_\infty$ following Eq.~\ref{eq:count_rate}.}
    \label{fig:HOM_methods}
\end{figure}

The HOM dip is more clearly revealed by projecting the 3D histrogram onto a 1D delay coordinate. In Method 1, the coincidence window is scanned across the four-fold events (dotted lines in Fig.~\ref{fig:HOM_methods}). To recover the full dip and the reference baseline \(R_\infty\), this approach requires window sizes significantly larger than the coherence time \cite{baghdasaryan2025}. In this approach, the baseline always includes contributions from both the TT and RR branches. Although Method 1 yields a high count rate for large windows, this comes at the cost of reduced visibility due to large coincidence windows in channels B and C.

In Method 2, the TT and RR branches are scanned separately (solid lines), allowing a meaningful HOM curve to be obtained even with narrow coincidence windows. The windows \(\Delta\tau_\text{AB}\) and \(\Delta\tau_\text{AC}\) must remain  sufficiently narrow relative to the separation of the two branches; otherwise, events from the opposite branch enter the integration window and artificially increase the apparent baseline. For large coincidence windows, the reference delay must therefore be chosen sufficiently far from the interference region that the TT and RR contributions are fully distinguishable. 

\begin{figure}[ht]
    \centering
    \includegraphics[width=\linewidth]{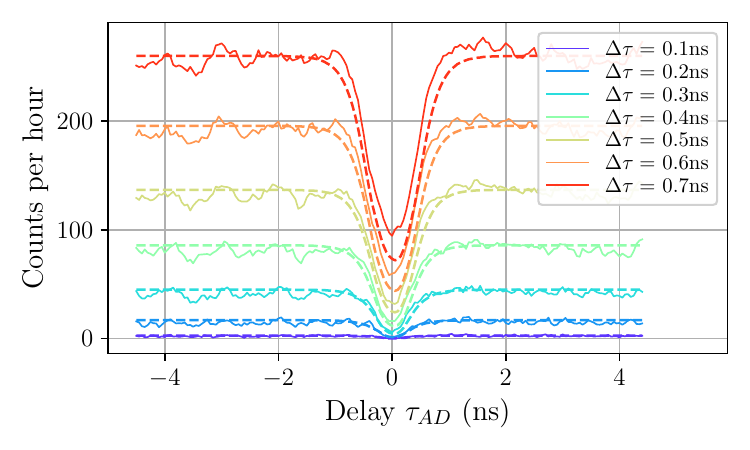}
    \caption{Comparison of the HOM curve obtained with Method 2 using experimental (solid) and simulated data (dashed) for different coincidence windows $\Delta\tau=\Delta\tau_\text{AB}=\Delta\tau_\text{AC}$. The bin width is 50\,ps. Results are shown for TT events; RR behaves analogously. As the coincidence window is increased, both the count rates outside and inside the dip rise, leading to a reduction in interference visibility. This effect is analyzed in more detail in Fig.~\ref{fig:cr} and Fig.~\ref{fig:vis}.}
    \label{fig:1d}
\end{figure}

Since narrow temporal post-selection in Method 2 provides higher interference visibility, the remainder of the discussion focuses on this method. For Method 2, we set equal concidence windwos ($\Delta\tau_\text{AB}=\Delta\tau_\text{AC}=\Delta\tau_\text{AD}=\Delta\tau$). The HOM curves obtained for different values of $\Delta\tau$ are shown in Fig.~\ref{fig:1d}. The evolution with the coincidence window width is in good agreement with the simulated data (dashed lines).

\subsection{Four-Fold Count Rates} \label{sec:count_rates}
\begin{figure}[ht!]
    \centering
    \includegraphics[width=\linewidth]{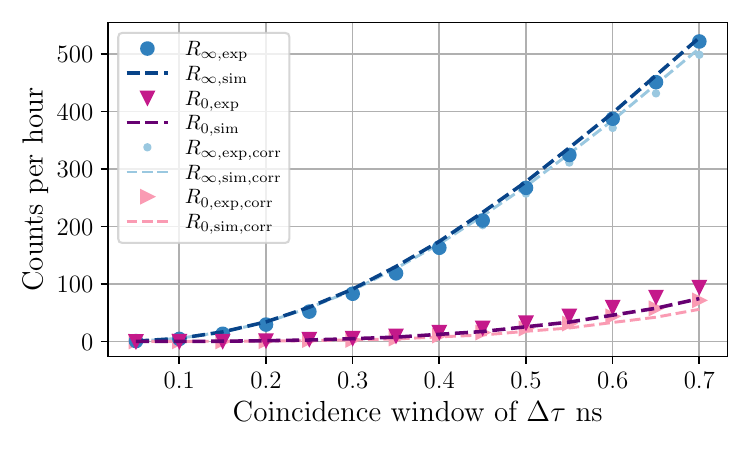}
    \caption{Total count rates at large delay $R_\infty=R_\text{TT}+R_\text{RR}$ (blue curves) and at zero delay $R_0$ (magenta curves). Both quantities increase with the size of the coincidence window. The simulated data (dashed lines) accurately reproduce the behaviour of $R_\infty$, but show slight deviations for the dip count rate $R_0$, due to the assumption of ideal interference and detection. Count rates corrected for the accidentals are shown in lighter colors.}
\label{fig:cr}
\end{figure}

The distinguishable rate of four-fold detections $R_\infty$ scales with the coincidence window, as shown in Fig.~\ref{fig:cr}, where all three windows are increased equally. The simulated count rate (blue dashed lines) agrees well with the experimental data (blue dots). The accidental-corrected count rates $R_{\infty, \text{corr}}$ (light blue) differ only marginally from the uncorrected count rate $R_{\infty}$ (dark blue in Fig.~\ref{fig:cr}), indicating that accidental coincidences contribute only a small fraction of the distinguishable reference rate.

The count rate at the interference minimum $R_0$, also increases with window width. Once the window exceeds the coherence time, it averages over the dip and reduces its depth, while also accumulating more accidental coincidences. The simulation reproduces the overall behaviour of $R_0$ with coincidence window,  but systematially underestimates the counts at the interference minimum. This is expected because the model (dashed line in Fig.~\ref{fig:cr}) does not account for higher-order interference or imperfect source purity. The model therefore provides an upper bound on achievable performance.

The accidental coincidence rate remains constant for different delay settings along the HOM curve. While its relative contribution is negligible at $R_\infty$, it becomes increasingly significant at the much lower count rate in the dip $R_0$. 

\subsection{Visibility}
The HOM interference contrast, or visibility, quantifies the suppression of coincidence events due to two-photon interference, and directly reflects the mode overlap of the interfering photons \cite{baghdasaryan2025, samara2019}. Since the visibility is determined from $R_\infty$ and $R_0$ (see Eq.~\eqref{eq:vis}), it is depends also to the choice of the coincidence window used to determine these quantities.

\begin{figure}[ht]
    \centering
    \includegraphics[width=\linewidth]{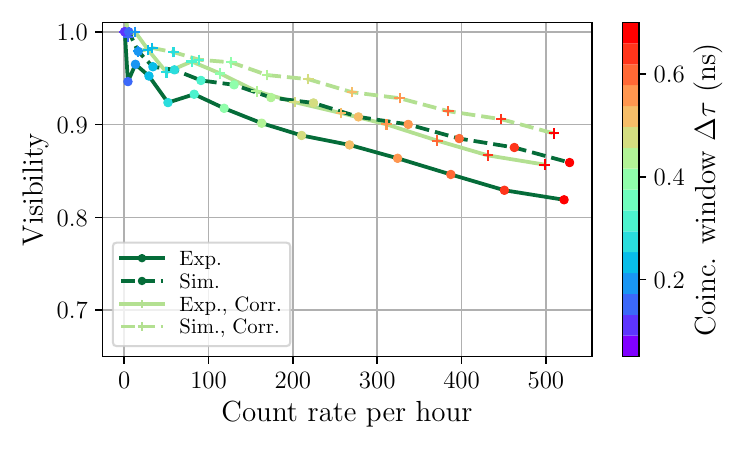}
    \caption{Trade-off between visibility and count rate. Different coincidence windows are indicated by colors according to Fig.~\ref{fig:1d}. The simulation represents an upper bound to ideal conditions. Visibilities corrected for the accidental coincidence events are shown in lighter green color.}
    \label{fig:vis}
\end{figure}

Fig.~\ref{fig:vis} shows the resulting trade-off between visibility and four-fold cout rate discussed in Sec. \ref{sec:count_rates}.  Increasing the coincidence window increases the number of four-fold detection events but also reduces visibility. Conversely, close to unit visibility can be achieved by drastically reducing the coincidence windows, but only at the expense of exceedingly low count rates. This clearly highlights that visibility alone is not a sufficient performance metric for experimental comparison, and that the count rate must be reported as well. 

For a given application, a minimum acceptable visibility can be set as a threshold, which in turn determines the maximum usable coincidence window with a given count rate. Choosing a coincidence window below the photon coherence time can improve the indistinguishability of the two sources by temporal filtering: selecting only the portion of the wave packet where the photons are fully indistinguishable and suppressing distinguishable contributions at the edges of the wave packet. For example, with $\Delta\tau=$600\,ps a count rate of 387 counts per hour is achieved in our experiment with a visibility of 84.6\,\%, while reducing $\Delta\tau$ to 300\,ps yields count rate of 83 counts per hour at an increased visibility of 93.2\,\%.

The simulation slightly underestimates the count rate at the interference dip, and therefore overestimates the visibility (dashed line in Fig.~\ref{fig:vis}). Several experimental imperfections can reduce the measured visibility below the simulated upper bound: i) partial distinguishability in polarization between the two photons reduces interference contrast directly; ii) thermal drift of the MRR resonance wavelengths or free spectral range can shift the spectral modes of the two sources apart, reducing their spectral indistinguishability; iii) timing jitter of the detection system reduces the effective temporal resolution, adding uncertainty to the localization of each photon within its wave packet. In the present experiment the jitter effect is small enough to be neglected, but it may need to be accounted for in experiments with shorter coherence times or longer jitter time; iv) higher-order photon pair emission, already discussed in the context of the dip count rate, contributes to additional accidental coincidences that further reduce the measured visibility. 

Since these effects are not included in the model, the simulated visibility represents an upper bound for an ideal source. Incorporating spectral drift, partial polarization distinguishability, and detection jitter would bring the simulation into closer agreement with the experimental results.

The model furthermore indicates that accidental four-fold events reduce visibility (light green curves in Fig.~\ref{fig:vis}). For example, a measured visibility of 93.2\,\% ($\Delta\tau=300$\,ps) corresponds to 96.8\,\% visiblity once accidentals are corrected. Improving heralding efficiency would suppress the accidental rate and thus increase measured visibility. Although this correction is useful for quantifying the impact of accidentals, it should not be used to adjust the performance of quantum key distribution or other protocols.

\section{Conclusion and Outlook}
\label{sec:conclusion}
We have demonstrated asynchronous four-fold HOM interference between photons generated by independent MRR sources and developed a quantitative model of the corresponding time-resolved four-fold coincidence distribution. By retaining all relative detection times and representing the events in a 3D delay space, the contributions from TT and RR photon-pair events, HOM visibility, as well as dominating types of accidentals, can be identified. We further analyzed how the size and placement of the coincidence windows affect the four-fold count rate and the interference visibility. The model, which requires only single-count rates and two-fold coincidence histograms as input, shows good agreement with the experimental data and provides an upper bound on the achievable performance even before the full four-photon experiment is conducted.

A central point of this work is that the methods discussed here are not different numerical ways of plotting the same HOM dip, but physically different temporal post-selection strategies. They differ in which detection channels are temporally filtered and in how the distinguishable reference $R_\infty$ is defined, and therefore select different subsets of four-fold events. Consequently, they yield different, but equally valid, visibility values. Even within a single method, the visibility can be traded against count rate over a wide range by adjusting the coincidence windows (Fig.~\ref{fig:vis}). A reported visibility is therefore only meaningful when stated together with the post-selection procedure, the coincidence windows, and the corresponding count rate. 

Looking ahead, the post-processing methods and associated trade-offs between quality and quantity of detection events are particularly relevant to quantum-network architectures in which photons from independent sources must interfere. CW operation is thereby particularly appealing as it avoids the need for a common optical clock or synchronized photon generation, shifting the temporal synchronization requirements from the photon wavepackets to time-resolved detection and timetag post-processing \cite{baghdasaryan2025}. Combined with integrated MRR sources that provide long photon coherence times and thereby relaxed temporal-matching requirements, this offers a promising route toward compact systems for efficient entanglement distribution and other interference-based quantum communication protocols.

\newpage
\appendix

\section{Biphoton mode function}\label{app:Theory_Biphoton_mode_fnc}
In the weak pumping regime, the quantum state of the photon pair generated in an MRR can be written as
\begin{equation}
\label{eq:ring_state_omega}
|\Psi\rangle
= \int d\omega_s\, d\omega_i\;
\Phi(\omega_s,\omega_i)\,
\ket{\omega_s,\omega_i}.
\end{equation}
Here $\Phi(\omega_s,\omega_i)$ is the biphoton mode function, also referred to as the Joint Spectral Amplitude (JSA). It encapsulates the spectral properties of the source, determined by the pump characteristics and the cavity density of states.

For SFWM in a MRR, two pump photons interact with the nonlinear medium and are converted into a pair of correlated signal and idler photons. The biphoton mode function can be written as the product of the pump spectral envelope and the cavity response functions,

\begin{equation}
\Phi(\omega_s,\omega_i)
\propto
\alpha(\omega_s+\omega_i)\,
L_s(\omega_s)\,L_i(\omega_i),
\end{equation}
where $\alpha(\omega)$ describes the pump spectrum and $L_\mu(\omega)$ is the cavity lineshape for modes $\mu\in\{s,i\}$. For a monochromatic CW pump at frequency $\omega_p$, energy conservation imposes $\omega_s+\omega_i = 2\omega_p$, so that

\begin{equation}
\Phi(\omega_s,\omega_i) \propto
\underbrace{\delta(\omega_s+\omega_i-2\omega_p)}_{\text{energy conservation}}
\;\times\;
\underbrace{L_s(\omega_s)\,L_i(\omega_i)}_{\text{cavity resonances}} .
\end{equation}

The cavity response is well described by a Lorentzian field amplitude,

\begin{equation}
L_\mu(\omega)
= \frac{1}{-i\bigl(\omega-\omega_{\mu,0}\bigr)+\gamma_\mu/2},
\end{equation}
where $\omega_{\mu,0}$ is the resonance frequency and $\gamma_\mu$ is the linewidth (full width at half maximum in angular frequency). Inserting this JSA into Eq.~\eqref{eq:ring_state_omega} and integrating over the delta function gives
\begin{equation}
|\Psi\rangle
= N \int d\omega_s\,
L_s(\omega_s)\,L_i(2\omega_p-\omega_s)\,
\ket{\omega_s,2\omega_p-\omega_s},
\end{equation}
with normalization factor $N$. It is convenient to introduce the detuning from the signal resonance,
$\Omega = \omega_s-\omega_{s,0}$, and to use the energy-conservation condition for the central frequencies
$2\omega_p = \omega_{s,0}+\omega_{i,0}$. Then, the state reads as
\begin{equation}
|\Psi\rangle
= N \int d\Omega\,
L_s(\omega_{s,0}+\Omega)\,L_i(\omega_{i,0}-\Omega)\,
\ket{\omega_{s,0}+\Omega,\omega_{i,0}-\Omega}.
\end{equation}
Defining the detuning basis states
$\ket{\Omega,-\Omega}\equiv\ket{\omega_{s,0}+\Omega,\omega_{i,0}-\Omega}$, we obtain

\begin{equation}
\label{eq:ringstate}
\ket{\Psi}
= N \int d\Omega\,\Phi^{\prime}(\Omega)\,\ket{\Omega,-\Omega},
\end{equation}
with the biphoton mode function
\begin{equation}
\Phi^{\prime}(\Omega)
= L_s(\omega_{s,0}+\Omega)\,L_i(\omega_{i,0}-\Omega)
= \frac{1}{-i\Omega+\gamma_s/2}\,\frac{1}{i\Omega+\gamma_i/2}.
\end{equation}
For equal linewidths $\gamma_s=\gamma_i=\gamma$ this simplifies to

\begin{equation}
\Phi^{\prime}(\Omega)
= \frac{1}{\Omega^2 + \bigl(\gamma/2\bigr)^2}.
\end{equation}

The temporal coherence function, $G^{(2)}(\tau)$, represents the probability density of detecting an idler photon at time delay $\tau$ relative to the signal photon. Theoretically, this is equivalent to the intensity of the temporal biphoton wavefunction $\Psi(\tau)$, which is the Fourier transform of the spectral amplitude $\Phi(\Omega)$:

\begin{equation}G^{(2)}(\tau) = \left| \Psi(\tau) \right|^2 \propto \left| \int_{-\infty}^{\infty} d\Omega , \Phi^{\prime}(\Omega) , e^{-i\Omega\tau} \right|\label{eq:G2}^2\end{equation}.

For the specific Lorentzian spectral function derived above, the Fourier transform yields a double-sided exponential decay:

\begin{equation}\Psi(\tau) \propto e^{-\frac{\gamma}{2}|\tau|}.\label{decay}\end{equation}
Squaring this amplitude gives the final temporal coherence function:

\begin{equation}G^{(2)}(\tau) \propto e^{-\gamma|\tau|}\label{coh}.
\end{equation}

This result indicates that the coincidence distribution decays exponentially with a time constant determined by the cavity lifetime ($1/\gamma$). In a realistic experiment, this function is convolved with the detectors timing jitter \cite{baghdasaryan2025}. The jitter can be modeled as a Gaussian convolution of the four-fold coincidence distribution, but for detectors with timing resolution much higher than the coherence time this effect can be neglected.

\section{Histogram Fit Function} \label{app:fit}
\begin{figure}[ht]
    \centering
    \includegraphics[width=0.75\linewidth]{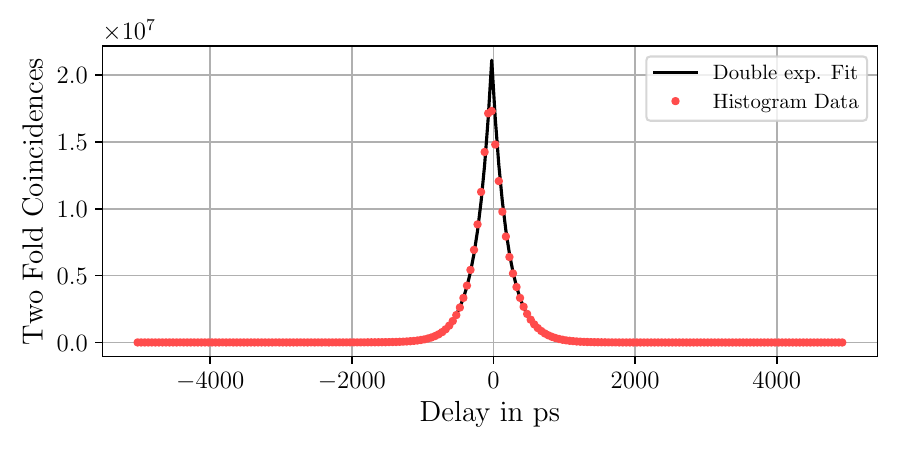}
    \caption{Experimental data of correlated two-fold events and the corresponding double-exponential fit, which is subsequently used to model the four-fold coincidence distribution. The integral overlap amounts to 98\,\%.}
    \label{fig:hist_2folds}
\end{figure}

In our model, we assume an ideal cavity characterized by a purely Lorentzian spectrum, which corresponds to a double-exponential temporal distribution, as explained in Appendix~\ref{app:Theory_Biphoton_mode_fnc}. In MRRs, however, random fluctuations and inhomogeneities typically lead to a Voigt spectral profile, i.e., the convolution of a Lorentzian with a Gaussian function. To keep the model simple, we neglect this Gaussian broadening. Instead, we fit a double-exponential function to the histogram, with the fit parameters optimized for maximal spectral overlap. In Fig.~\ref{fig:hist_2folds}, we show this fit for the histogram of channels A and B, while analogous results are obtained for the other channel combinations. The resulting fit parameters are $R_\text{AB} =1490$ counts per second for the center bin and a decay time $1/\gamma= 216$ ps, achieving an integral overlap of 98\,\% with the measured histogram.

\section{Noise distribution} \label{app:noise}
\begin{figure}[ht]
    \centering
    \includegraphics[width=0.95\linewidth,trim={0 0 0 0},clip]{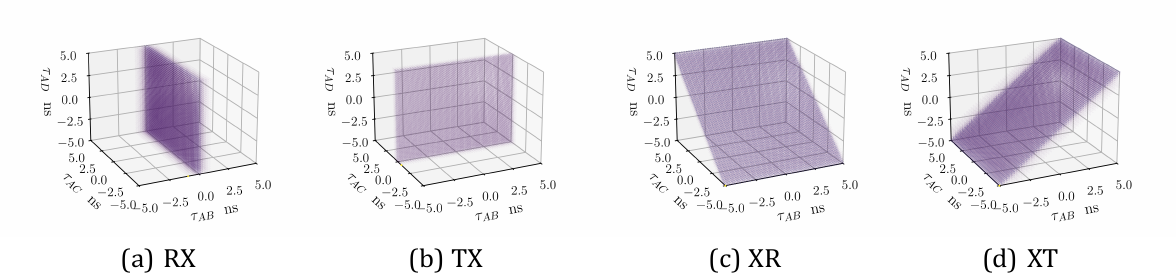}
        \caption{Main contributions to accidental events arising from the combination of a correlated and an uncorrelated photon pair, leading to a four-fold coincidence. (a) The correlated photon pair is detected in channels A and B, whereas two uncorrelated events are detected in the other two channels, D and C, as defined for $r_\text{RX}$ in Eq.~\eqref{eq:acc}. (b) The correlated pair is detected in channels A and C, as defined for $r_\text{TX}$. (c) The correlated pair detected in channels D and C, as defined for $r_\text{XR}$. (d) The correlated pair detected in channels D and B, as defined for $r_\text{XT}$.}
    \label{fig:noise_planes}
\end{figure}

In the 3D histogram (Fig.~\ref{fig:3dHist}), accidental counts appear primarily along central planes where the delay between two of the three channels is zero. This indicates that most accidental events are caused by a photon pair, one photon in an interacting channel (B or C) correlated with a photon in a heralding channel (A or D), while the other two channel combinations contain uncorrelated photons. By modeling the contribution of true four-fold correlation events and accidental events, these contributions can be evaluated independently. The accidental events, defined in Eq. \eqref{eq:acc}, are distributed in planes of the 3D histogram. In the case of finding the correlated photon in channels A (the heralding and reference channel) and B (one of the interacting channels), the plane of accidental events is vertical, located at $\tau_\text{AB}=0$ (as defined for $r_\text{RX}$ in Eq.~\ref{eq:acc}, see Fig.~\ref{fig:noise_planes}a). When the pair is detected in channels A and C, the plane is located at $\tau_\text{AC}=0$ (as defined for $r_\text{TX}$, see Fig.~\ref{fig:noise_planes}b). In the case of correlated photon pair detections in channel B and heralding channel D (as defined for $r_\text{XT}$, see Fig.~\ref{fig:noise_planes}c), the accidentals are distributed over a diagonal plane at $\tau_\text{AB}=\tau_\text{AD}$, whereas for the combination C and D (as defined for $r_\text{XR}$, see Fig.~\ref{fig:noise_planes}d), the plane is found at $\tau_\text{AC}=\tau_\text{AD}$.

\begin{figure}[ht]
    \centering
    \includegraphics[width=0.4\linewidth]{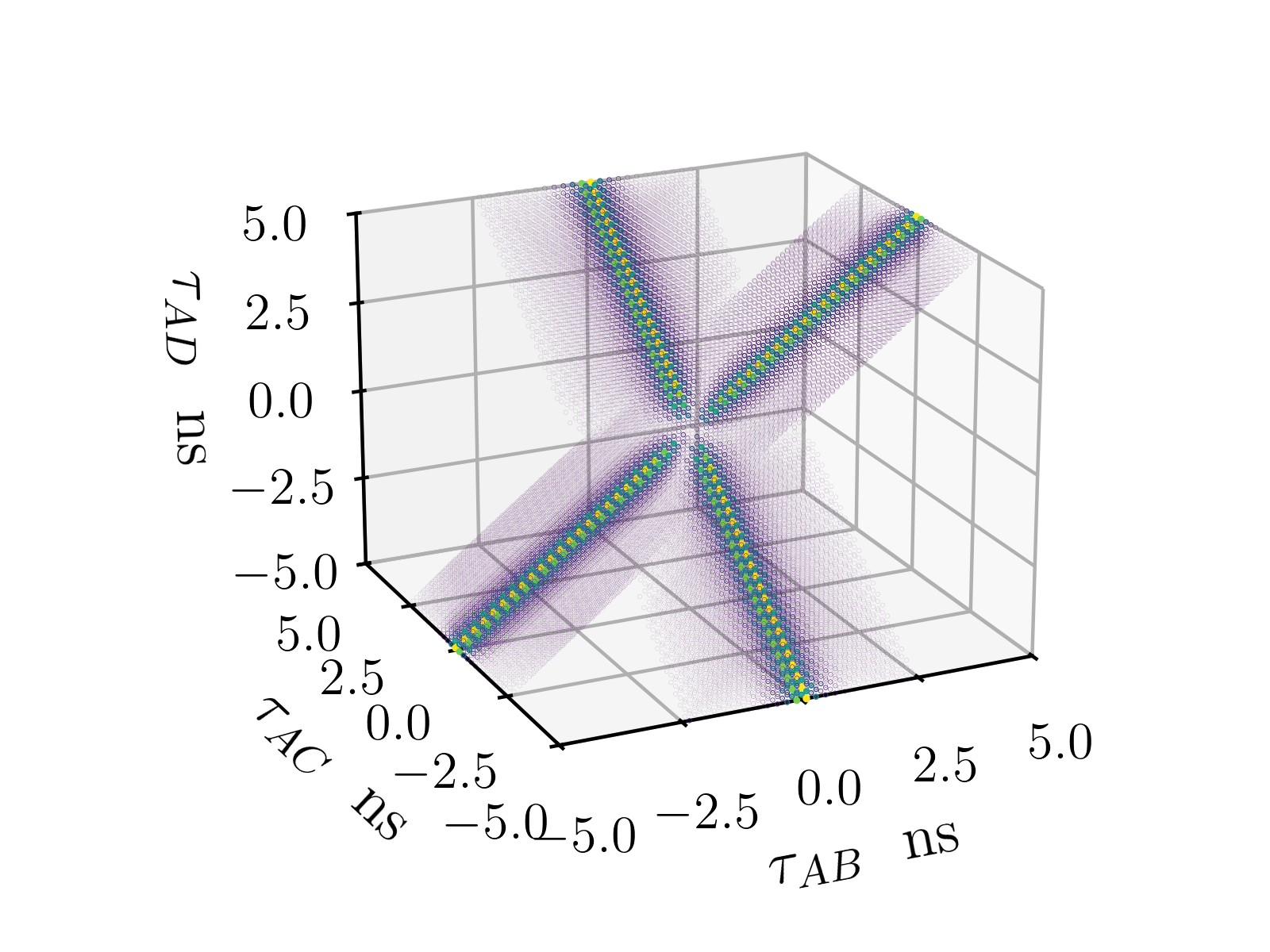}
    \caption{True four-folds shown without considering accidental events. The four-folds are distributed along two branches: along the diagonal $\tau_\text{AD}=\tau_\text{AB}$ for TT events and along $\tau_\text{AD}=\tau_\text{AC}$ for RR events}
    \label{fig:rr_tt}
\end{figure}

Additionally, the true four-fold events $r_ \text{HOM}$ can be visualized without the background of accidental events, as shown in Fig.~\ref{fig:rr_tt}, where the two crossing branches correspond to TT and RR events.

\section{Alternative Method to obtain HOM Curves with Detailed Analysis of Coincidence Window Size} \label{app:HOM_method_3}
\begin{figure}[ht]
    \centering
    \includegraphics[width=0.95\linewidth,trim={0 0 0cm 0},clip]{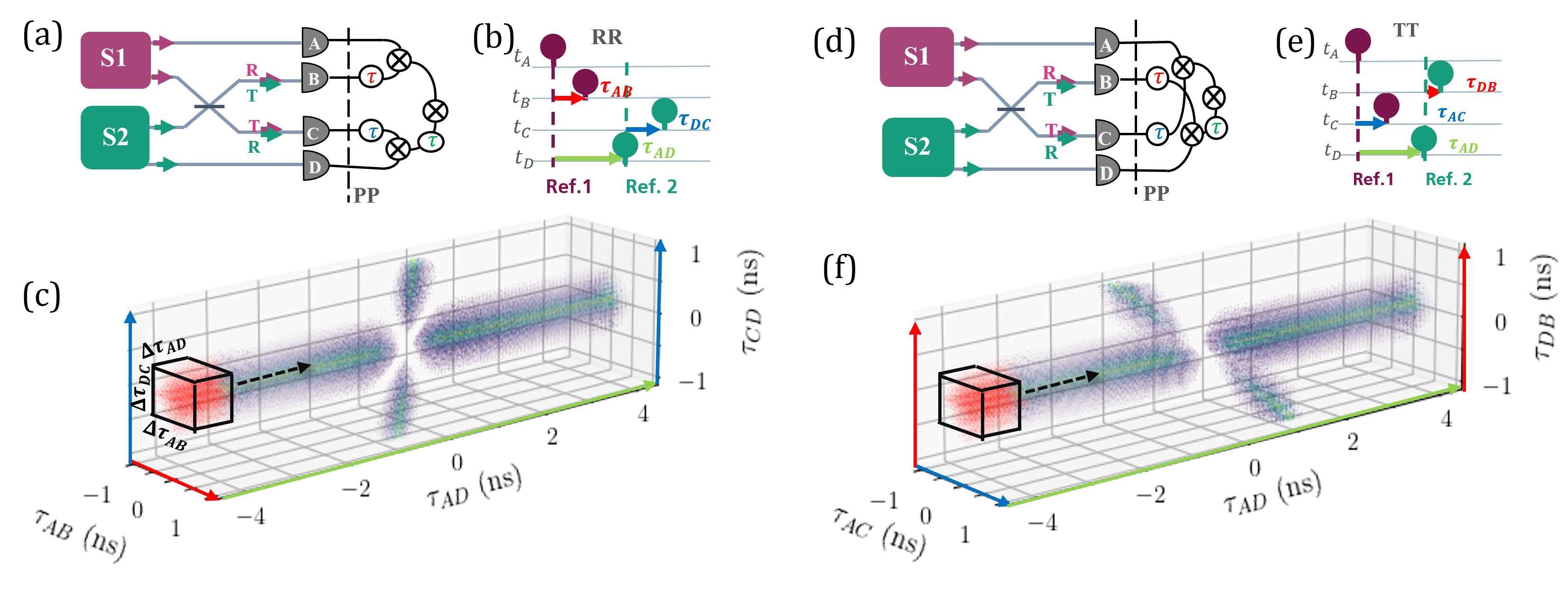}
    \caption{Schematics of the newly defined temporal axes (upper row) and modeled four-fold data shown in 3D histograms (lower row): (a) time-delay configuration for each channel to obtain the RR events; (b) representation of the four time-tagged channels, with channels A and D serving as time references for channels B and C, respectively; (c) 3D histogram with coordinates defined to align with the RR events; (d) time-delay configuration for each channel to obtain the TT events, (e) representation of the four time-tagged channels, with channels A and D serving as time references for channels C and B, respectively; (f) 3D histogram with coordinates defined to align with the TT events.}
    \label{fig:3d_2d_1d}
\end{figure}

To independently analyze the coherence properties of each source directly in the 3D histogram, we can apply a coordinate transformation. For this, we define one axis as the relative delay of two-fold events from source S1, a second axis as the relative delay of two-fold events from source S2, and a third axis as the delay between the heralding events. Thus, for the observation of RR events, we set the axes to $\tau_\text{AB}= t_\text{B}-t_\text{A}$, $\tau_\text{DC}= t_\text{C}-t_\text{D}$ and $\tau_\text{AD} = t_\text{D}-t_\text{A}$. In this representation, channel D is used as the reference time for its correlated event in channel C, such that the corresponding counts are no longer distributed along a diagonal but aligned with a coordinate axis. For the case of TT events, we set the coordinates analogously to $\tau_\text{AC}= t_\text{C}-t_\text{A}$, $\tau_\text{DB}= t_\text{B}-t_\text{D}$ and $\tau_\text{AD} = t_\text{D}-t_\text{A}$ (Fig.~\ref{fig:3d_2d_1d}). These transformations enable an independent investigation of RR and TT events.

\begin{figure}[ht]
    \centering
    \includegraphics[width=0.48\linewidth]{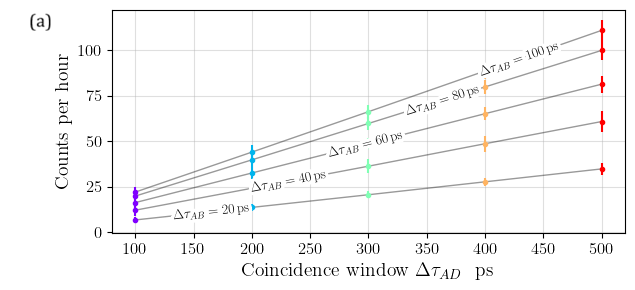}
    \includegraphics[width=0.48\linewidth]{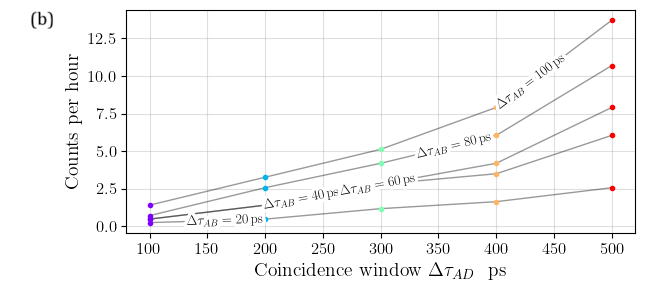}  
    \caption{By fixing the coincidence window $\Delta\tau_\text{AC}=300$\,ps, (a) the count rate at a delay $\tau_\text{AD}\gg 0$ grows linearly for increasing $\Delta \tau_\text{AD}$, (b) while the count rate at the dip at $\tau_\text{AD}= 0$ increases with a higher-order dependence.}
    \label{fig:CR_M3}
\end{figure}

The trade-off between visibility and count rate can be controlled by varying the coincidence window sizes. By fixing the coincidence window in the $\tau_\text{AB}$ and $\tau_\text{AC}$ dimension to 300\,ps, thus constraining the relative time between the signal and idler photons within each pair, the count rate scales linearly with the window size $\Delta\tau_\text{AD}$ (Fig.~\ref{fig:CR_M3}). In contrast, increasing $\Delta\tau_\text{AB}$ and $\Delta\tau_\text{AC}$ acts as a windowing function that averages over the delay. This reduces the resolution of the HOM dip and consequently, the count rate at the dip increases and the visibility decreases.

\begin{figure}[ht]
    \centering
    \includegraphics[width=0.48\linewidth]{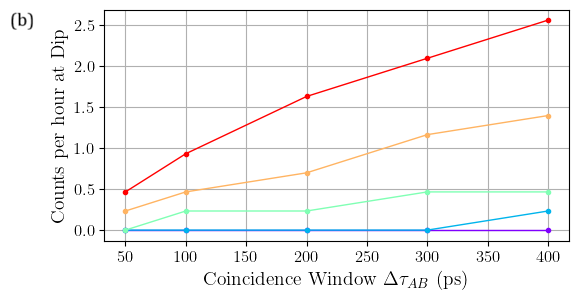}
    \includegraphics[width=0.48\linewidth]{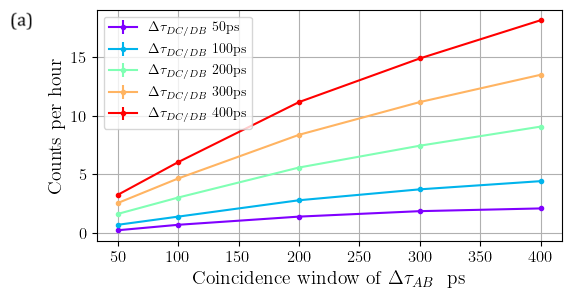} 
    \caption{For a fixed coincidence window $\Delta\tau_\text{AD}=300$\,ps and under the condition $\Delta\tau_\text{DC}=\Delta\tau_\text{DB}$: (a) the count rate at a delay depends on the coherence function $G^{(2)}$, while (b) the count rate at the dip integrates more accidental events as $\Delta \tau_\text{AB}$ increases.}
    \label{fig:CR_M3_2}
\end{figure}

If instead the window size $\Delta\tau_\text{AD}$ is fixed to  300\,ps and either $\Delta\tau_\text{AB}$ ($\Delta\tau_\text{AC}$) or $\Delta\tau_\text{DC}$ ($\Delta\tau_\text{DB}$) is varied along the RR (or TT) branch, the count rate increases as $\propto1-G^{(2)}$ for both cases (Fig.~\ref{fig:CR_M3_2}). When the coincidence window is much larger than the coherence time, it mainly integrates more accidental coincidences rather than increasing the true four-fold count rate. As the window widens, the counts at the dip rise because more events from the perpendicular branch (TT or RR) fall within the window, thereby reducing the visibility. Unlike approaches based on a fixed coincidence window, this trade-off can be flexibly adjusted to suit the experimental requirements. 

However, this method does not allow for simply summing the count rates obtained separately for RR and TT contributions, as in Eq.~\ref{eq:count_rate}, because the two quantities are obtained after different coordinate transformations. Therefore, the visibility cannot be determined straightforwardly, since it requires the reference value $R_\infty$, which should include both RR and TT contributions. A direct calculation without taking into account the coordinate transformations is not comparable to the visibilities obtained using Method 1 and Method 2 in the main text, as described in Sec.~\ref{sec:methods}.

\begin{backmatter}
\bmsection{Funding}
This work was financially supported by the Federal Ministry of Research, Technology and Space (BMFTR) through the QuNET initiative, BasiQ and the Carl-Zeiss-Stiftung within the Carl-Zeiss-Stiftung Center for Quantum Photonics (CZS QPhoton) under the project ID P2021-00-019.

\bmsection{Acknowledgment}
The presented results were partially acquired using facilities and devices funded by the Free State of Thuringia within the application center Quantum Engineering.
S.H., M.C.P. and F.S. are part of the Max Planck School of Photonics supported by Dieter Schwarz Foundation, Max Planck Society and Fraunhofer Society. 

\bmsection{Disclosures}
The authors declare no conflicts of interest.

\bmsection{Data availability} Data underlying the results presented in this paper are not publicly available at this time but may be obtained from the authors upon reasonable request.

\bmsection{Author Contributions}
S.H. -- Conceptualization, Methodology, Software, Formal Analysis, Data Curation, Writing – Original Draft, Visualization;
M.C.P. -- Conceptualization, Supervision, Writing - Review \& Editing, Project Administration;
A.M.T. -- Conceptualization, Investigation (Conducting experiment), Software, Writing – Review \& Editing;
B.B. -- Validation, Formal Analysis, Writing – Review \& Editing;
K.L.M. -- Conceptualization,  Writing – Review \& Editing;
F.S. -- Conceptualization, Supervision, Project Administration, Resources, Writing – Review \& Editing, Funding Acquisition

\end{backmatter}


\bibliography{references}






\end{document}